\documentclass[aps,prl,groupedaddress,floatfix,twocolumn]{revtex4-1}

\usepackage{graphicx,
            caption,
            subcaption,
            amsmath,
            braket,
            ragged2e,
            xcolor,
            xspace,
            nicefrac,
            lipsum,
            nameref,
            textcomp,
            urwchancal,
            upgreek,
            isotope,
            hyperref,
            cleveref,
            siunitx
            }

\DeclareCaptionLabelSeparator{dot}{. }
\makeatletter
\def\justified{
	\let\\\@normalcr
	\@rightskip\z@skip \rightskip\@rightskip
	\leftskip\z@skip
	\parindent 0em\relax
	\setlength{\parfillskip}{0pt plus 1fil}}
\DeclareCaptionJustification{justified}{\justified}
\hypersetup{colorlinks=true, linkcolor=blue,citecolor=blue,urlcolor=blue}
\def\unit #1 #2 {\SI{#1}{#2}\xspace}
\def\range #1 #2 #3 {\SIrange{#1}{#2}{#3}\xspace}
\DeclareSIUnit\gauss{G}

\newcommand{\myref}[2][]{Fig.~\hyperref[#2]{\ref*{#2}#1}}
\newcommand{\Myref}[2][]{Figure~\hyperref[#2]{\ref*{#2}#1}}
\newcommand{\Mytabref}[2][]{Table~\hyperref[#2]{\ref*{#2}#1}}
\newcommand{\setlabel}[1]{\edef\@currentlabel{#1}\label}

\begin{document}

\title{Two-photon cooling of calcium atoms}

\date{\today}
	
\author{Wojciech Adamczyk}
\thanks{These authors contributed equally.}
\author{Silvan Koch}
\thanks{These authors contributed equally.}
\author{Claudia Politi}
\author{Henry Fernandes Passagem}
\author{\mbox{Christoph Fischer}}
\author{Pavel Filippov}
\author{Florence Berterottière}
\author{Daniel Kienzler}
\author{Jonathan Home}
\thanks{Contact author: \mbox{\url{jhome@phys.ethz.ch}}}

\affiliation{
    Institute for Quantum Electronics, ETH Z\"urich, Otto-Stern-Weg 1, 8093 Z\"urich, Switzerland\\
    and Quantum Center, ETH Z\"urich, 8093 Z\"urich, Switzerland
}

\begin{abstract}
     We demonstrate sub-Doppler cooling of calcium atoms using a two-photon transition from the ${^1}S_0$ ground state to the upper $4s5s~{^1}S_0$ state via the ${^1}P_1$ intermediate state. We achieve temperatures as low as $\SI{260}{\micro\kelvin}$ in a magneto-optical trap (MOT), well below the Doppler limit ($T_{\text{D}} = \SI{0.8}{\milli\kelvin}$) of the ${^1}P_1$ state. We characterize temperature, lifetime and confinement of the MOT over a range of experimental parameters, observing no reduction in lifetime due to coupling to the higher state. We perform theoretical simulations of the cooling scheme and observe good agreement with the experimental results. The two-photon cooling scheme presented in this work provides an alternative to the standard Doppler cooling applied to alkaline-earth atoms, based on a sequence of two magneto-optical traps. The advantages of our scheme are the possibility of varying the effective linewidth of the ${^1}P_1$ state, a higher transfer efficiency (close to 100$\%$), and a more straightforward experimental implementation.
\end{abstract}

\maketitle

\section{I. Introduction}\label{sec:introduction}
 
Alkaline earth(-like) atoms, characterized by two valence electrons, offer transitions with linewidths ranging from tens of $\SI{}{\mega\hertz}$ down to below $\SI{}{\milli\hertz}$. This broad range has enabled breakthrough results in optical atomic clocks~\cite{Ludlow2015Optical,Bothwell2019JILA}, and stimulated their use to probe effects beyond the standard model~\cite{Tiberi2024Searching}. In the context of quantum simulation and computation, arrays of individual alkaline-earth(-like) atoms have emerged as a promising platform, where, in addition to their scalability, the long-lived metastable states can be used for motional ground state cooling~\cite{Cooper2018alkaline,Norcia2018microscopic,Hoelzl2023motional}, qubit readout and manipulation~\cite{Saskin2019narrow-line,Ma2022universal}, and providing access to single-photon Rydberg excitation~\cite{Madjarov2020High} and quantum erasure conversion~\cite{Wu2022Erasure,Scholl2023Erasure,Ma2023High}. The presence of a second valence electron provides an additional degree of control over both the external and internal degrees of freedom of the atoms. For instance, in atoms excited to circular Rydberg states, the inner electron can assist in trapping~\cite{Wilson2022Trapping,Hoelzl2024long-lived}, cooling~\cite{Bouillon2024Direct,Lachaud2024slowing} and manipulating the state of the outer electron~\cite{Muni2022Optical,Burgers2022Controlling,Wirth2024coherent}.

Many experiments performed with neutral atoms involve initial cooling and trapping of atoms in a magneto-optical trap (MOT), with subsequent transfer to optical dipole traps, tweezers or lattices. The initial cooling step is required because the optical traps have trap depths of around \SI{1}{\milli\kelvin} that are far below room temperature. Therefore, starting with atomic samples with temperatures on the order of tens of microkelvin is beneficial. Alkali atoms, with hyperfine levels in the ground state, benefit from sub-Doppler cooling schemes, such as polarization gradient cooling~\cite{Lett1988Observation,Dalibard1989Lasercooling}. Bosonic isotopes of alkaline-earth atoms, while possessing excellent properties for atomic clocks, have no hyperfine structure and no degeneracy in the ground state. Thus, no natural sub-Doppler cooling scheme exists. To overcome this challenge, two stages of magneto-optical trap are typically applied to reach microkelvin temperatures. The first stage operates on the broad dipole-allowed transition from the ground state, offering a large capture velocity for initial trapping. A second stage is then often employed, which uses a narrower linewidth transition to attain lower temperatures. The transfer of cold atoms from one stage to the other increases the complexity of the experiment and leads to atom losses. For atoms such as calcium, where the narrow transition results in forces comparable to gravity, other schemes such as quenched cooling~\cite{Binnewies2001doppler, Curtis2001Quenched, Curtis2003Quenched} need to be implemented to effectively broaden the linewidth of the narrow state. This conflicts with an additional challenge of alkaline-earth atoms, which is the presence of excited metastable states, leading to leakage channels from the cooling transition. Alternatively, calcium atoms can be cooled to sub-Doppler temperatures by simultaneously operating two distinct magneto-optical traps: one based on the broad dipole-allowed transition and another one on the closed transition $4s4p~{^3}P_2 \rightarrow 4s4d~{^3}D_3$ ~\cite{Gruenert2002subdoppler}. In this approach, atoms are initially captured and cooled in the standard MOT. They then decay to the metastable $4s4p~{^3}P_2$ state, where they are recaptured and further cooled by the second MOT.

In this work, we cool $^{40}$Ca atoms to sub-Doppler temperatures by implementing the two-photon cooling scheme proposed in Ref.~\cite{Magno2003two-photon} for alkaline-earth atoms. A similar scheme was experimentally realized with magnesium atoms in Refs.~\cite{Malossi2005two-photon,Mehlstaeubler2008observation}, by coupling the $3s3p~{^1}P_1$ to the $3s3d~{^1}D_2$ state. However, in calcium, the $4s4d~{^1}D_2$ state suffers from atom losses due to leakage channels to the triplet states, limiting the cooling time. To overcome this issue, we couple the ground state ${^1}S_0$ to the upper $4s5s~{^1}S_0$ via the intermediate ${^1}P_1$, realizing a cooling scheme which is less prone to losses. We achieve temperatures as low as $\SI{260}{\micro\kelvin}$ in a magneto-optical trap and show a transfer efficiency close to 100$\%$ from the standard MOT operating on the broad transition. Finally, we verify that the atomic lifetime is not limited by the two-photon cooling scheme. 
 
\begin{figure*}
    \centering
    \includegraphics[width=\textwidth]{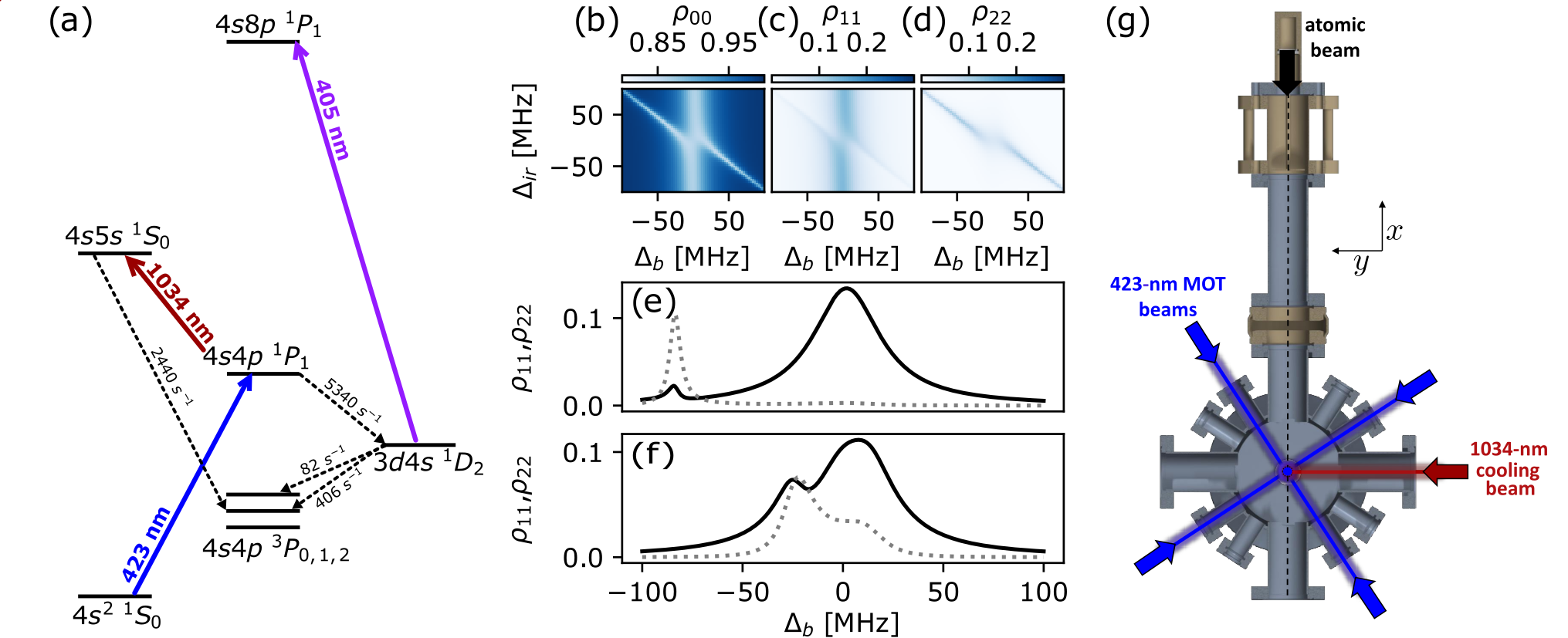}
    \caption{{\bf Calcium level scheme, atomic population, experimental apparatus.} (a) Relevant energy levels of calcium. The blue and red arrows indicate the transition used for the magneto-optical trap and the two-photon cooling, respectively. The black dashed arrows denote the decay channels: from the $4s5s~{^1}S_0$ state to the ${^3}P_1$ state with a rate of $\SI{2440}{\second^{-1}}$, from the $4s4p~{^1}P_1$ state to the $3d4s~{^1}D_2$ state with a rate of $\SI{5340}{\second^{-1}}$, and from the $3d4s~{^1}D_2$ state to the ${^3}P_1$ and ${^3}P_2$ with a rate of $\SI{406}{\second^{-1}}$ and $\SI{82}{\second^{-1}}$, respectively~\cite{Mills2017efficient}. The purple arrow shows the transition used to repump the atoms from the ${^1}D_2$ state into the cooling cycle. (b-f) Theoretical calculation of the population in the ${^1}S_0$ (b), ${^1}P_1$ (c), and $4s5s {^1}S_0$ (d) state, as a function of 423-nm detuning $\Delta_{\text{b}}$, and 1034-nm detuning, $\Delta_{\text{ir}}$. These were calculated with the experimental parameters: $I_{\text{b}} = 0.06I_{\text{sat,b}}$, $I_{\text{ir}} = 100I_{\text{sat,ir}}$, where $I_{\text{sat,b}} = \SI{60}{\milli\watt\per\centi\meter\squared}$ and $I_{\text{sat,ir}} = \SI{0.46}{\milli\watt\per\centi\meter\squared}$ are the saturation intensities of the 423-nm and 1034-nm transitions, respectively. Population of the ${^1}P_1$ state (black line), and $4s5s~{^1}S_0$ state (gray dashed line) for $\Delta_{\text{ir}} = \SI{80}{\mega\hertz}$ (e), and $\Delta_{\text{ir}} = \SI{11.8}{\mega\hertz}$ (f). (g) Sketch of the experimental vacuum chamber. The atomic flux is generated by a home-made oven, and reaches the main chamber, where the atoms are cooled and trapped in a magneto optical trap on the 423-nm transition (blue beams). A 1034-nm laser (red beam) is used to perform two photon cooling. Only the horizontal beams are shown in the Figure. The 405-nm repumper is superimposed with the horizontal MOT beams. A resonant 423-nm beam (not shown) overlapped with the 1034-nm beam is shone onto the atoms to perform absorption imaging.}
	  \label{fig:1} 
\end{figure*}


\section{II. Two-Photon Cooling Scheme Overview}\label{sec:overview}

For alkaline-earth atoms, initial Doppler cooling typically happens on the ${^1}S_0 \rightarrow {^1}P_1$ transition, resulting in a Doppler temperature limit on the order of a millikelvin~\cite{Xu2002Dynamics}. A common second step involves Doppler cooling on the spin-forbidden ${^1}S_0 \rightarrow {^3}P_1$ transition. For magnesium, calcium, strontium, and ytterbium the linewidths of these transitions are $\Gamma/2\pi = \SI{30}{\hertz}, \SI{370}{\hertz}, \SI{7.5}{\kilo\hertz}$, and $\SI{180}{\kilo\hertz}$~\cite{Rehbein2007Optical, Ludlow2015Optical}, respectively. Even for the largest of these, the big discrepancy between the linewidths of the first and second-stage transitions leads to low transfer efficiency, which can be extended up to 50$\%$ by spectrally broadening the laser light used to drive the narrow transition~\cite{Katori1999Magneto-Optical,Chanelière2008Three}, or by implementing more sophisticated schemes, such as a SWAP MOT~\cite{Norcia2018narrow-line,Snigirev2019Fast}. While such an approach works for strontium, it is impractical for species with even narrower second stage transitions, such as calcium and magnesium, where the resulting cooling force is comparable to gravity. The complexity related to the implementation of two magneto-optical traps and the low transfer efficiency from the broad to the narrow MOT makes it favorable to examine alternative solutions.

We investigate an approach in which the first step of cooling on the regular broad transition is then modified with a dressing beam tuned to the ${^1}P_1 \rightarrow nsn's~{^1}S_0$ transition, allowing an effective two-photon Doppler cooling~\cite{Magno2003two-photon}, where the temperature limit is now set by the linewidth of the upper ${^1}S_0$ state. The linewidths of the relevant upper states for magnesium, calcium, strontium, and ytterbium are $\Gamma/2\pi = \SI{3.53}{\mega\hertz}, \SI{3.88}{\mega\hertz}, \SI{2.96}{\mega\hertz}, \SI{3.50}{\mega\hertz}$~\cite{Magno2021Laser}. The corresponding Doppler temperatures are $\SI{85}{\micro\kelvin}$, $\SI{93}{\micro\kelvin}$, $\SI{71}{\micro\kelvin}$, $\SI{84}{\micro\kelvin}$, respectively. We note that for calcium the linewidth of the upper ${^1}S_0$ state is not well known, with several works reporting different values~\cite{Magno2003two-photon, Mills2017efficient}. This two-photon cooling scheme allows for a near 100$\%$ transfer efficiency and constitutes a simpler implementation compared to the narrow-line cooling mentioned before. The exact orientation of the dressing laser beam is not critical, since it is used to modify the population of the ${^1}P_1$ state. Moreover, the relatively small mismatch of linewidths between the dressed and standard MOT enables operation with a static magnetic field, and does not require any additional broadening techniques to enhance transfer efficiency.

This cooling method is not restricted to the ${^1}P_1 \rightarrow {^1}S_0$ transition, but can be realized by coupling the ${^1}P_1$ energy level to any state. For instance, a similar approach was demonstrated with magnesium atoms by coupling the ${^1}P_1$ state to the narrower ${^1}D_2$ state~\cite{Malossi2005two-photon,Mehlstaeubler2008observation}. However, for this scheme to be effective, a few requirements need to be met: (1) The linewidth should be sufficiently narrow to reach the target temperature; (2) The upper state should not have unwanted decay channels to any long-lived states that are not trapped by the MOT light; (3) For single beam use, the recoil of the secondary photon needs to be smaller than the recoil of the main cooling transition. The two-photon cooling scheme realized in our work applies to a regime in which the scheme can be treated as an effective two-photon Doppler cooling on the upper transition. For smaller detunings from the intermediate state, a different cooling regime can be realized in which the coherent coupling of the three states gives rise to an electromagnetic-induced transparency (EIT) feature, see Refs.~\cite{Morigi2007two-photon,Dunn2007coherent}.

In the first stage of cooling, calcium atoms are cooled by coupling the $4s^2~{^1}S_0$ ground state to the upper state $4s4p~{^1}P_1$ via the strong 423-nm dipole-allowed transition with a linewidth $\Gamma_{\text{b}}/2\pi = \SI{35}{\mega\hertz}$. Figure~\ref{fig:1}(a) depicts the energy level scheme with the most relevant transitions. As in other alkaline-earth atoms, the $4s4p~{^1}P_1$ state is weakly coupled to the $4s4p~{^3}P_{0,1,2}$ state through the $3d4s~{^1}D_2$ state. For calcium, the ${^3}P_{2}$ state is a metastable state with a lifetime of $\SI{118}{\minute}$~\cite{Derevianko2001Feasibility,Mills2017efficient}. In this state, the atoms are not subjected to the cooling light and escape the MOT region, leading to atom losses. To overcome this issue, a repump laser can be used to couple the ${^1}D_2$ to upper states, allowing the atoms to decay back to the ground-state and re-enter into the cooling cycle. In our experiment, we couple the ${^1}D_2$ state to the $4s8p~{^1}P_1$ state with a wavelength of $\SI{405}{\nano\meter}$, which leads to an improvement in the MOT atom number of a factor of 20. Other repumping schemes for calcium atoms are described in Ref.~\cite{Mills2017efficient}.   

The presence of a dressing beam, which couples the ${^1}P_1$ state to the $4s5s~{^1}S_0$ state via the 1034-nm transition (see Fig.~\ref{fig:1}(a)), alters the linewidth profile of the ${^1}P_1$ transition. Theoretical calculations of the dressing beam's effect are shown in Figs.~\ref{fig:1}(b-f). The calculated populations as a function of detunings ($\Delta_{\text{b}}$, $\Delta_{\text{ir}}$) for the energy levels ${^1}S_0$, ${^1}P_1$, and $4s5s{^1}S_0$ are presented in Figs.~\ref{fig:1}(b,c,d), respectively. The populations are calculated for typical experimental parameters (for more details about the theoretical model see Appendix\,\ref{AppendixA}). The plots display the effect of the 1034-nm laser beam on the populations, with a distinctive diagonal feature close to the two-photon resonance. Close to these conditions the population in the ${^1}P_1$ state comes from a decay from the excited ${^1}S_0$ energy level. 

If the primary cooling beam is close to resonance, the dressing laser beam gives rise to an EIT feature and, when far off resonance, the three-level system can be treated as an effective two-level system. In both cases, the presence of the dressing beam creates a narrow feature in the absorption spectrum of the primary laser. This feature has an asymmetric Fano line shape~\cite{Dunn2007coherent} and its width approaches the linewidth of the secondary transition. Figures~\ref{fig:1}(e,f) present these two regimes, where the detuning of the secondary beam is set to $\Delta_{1034} = \SI{80}{\mega\hertz}$ and $\Delta_{1034} = \SI{11.8}{\mega\hertz}$, respectively.


\section{III. Experiment}
\setlabel{III}{sec:experiment}

A schematic of the beam layout is given in  Fig.~\ref{fig:1}(g). The atomic beam, generated from a home-made effusive oven at about $\SI{600}{\degreeCelsius}$, propagates along the $x$-axis. The atoms are captured in a magneto-optical trap operating on the 423-nm transition (${^1}S_0 \rightarrow {^1}P_1$), with three pairs of counter-propagating laser beams. One of these pairs propagates along the gravity direction ($z$-axis), and two in the horizontal $x-y$ plane. The two horizontal beams propagate with an angle of 30$^\circ$ and -60$^\circ$ with respect to the $x$-axis. We perform the two photon cooling by adding a single 1034-nm laser beam propagating along the $y$ axis. The atomic cloud is imaged using absorption imaging by shining a resonant 423-nm laser beam along the $y$-axis.

We typically start by producing a MOT of around 1$\times 10^6$ atoms with a temperature of $\SI{4}{\milli\kelvin}$ at a detuning of $\Delta_{\text{b,MOT}} \approx - 2\Gamma_{\text{b}}$. The magnetic field gradient in the $x-y$ plane is set to \SI{135}{\gauss\per\centi\meter}. The MOT beams have a waist of $\SI{1.8}{\milli\meter}$ with a total power of $\SI{20}{\milli\watt}$, resulting in an intensity of $I_{\text{b,MOT}} = 3.8I_{\text{sat,b}}$, where $I_{\text{sat,b}} = \SI{60}{\milli\watt\per\centi\meter\squared}$ is the saturation intensity of the 423-nm transition. The 423-nm light is frequency stabilized to the atomic resonance via polarization spectroscopy~\cite{Wieman1976Doppler-free,Shimada2013Asimplified}, and is delivered to the experiment after passing through an acousto-optic modulator (AOM) in a double-pass configuration. The AOM is used to control the intensity and frequency of the light.

The two-photon cooling sequence starts with loading the MOT for \SI{3}{\second}. We then apply a \SI{2}{\milli \second}-long intermediate step, referred to as pre-cooling. In this step, the total power of the MOT light is reduced to $\SI{0.6}{\milli\watt}$ ($I_{\text{b,pc}} = 0.1I_{\text{sat,b}}$) and the detuning reduced to $\Delta_{\text{b,pc}} = -  \Gamma_{\text{b}}$. This decreases the temperature of the cloud to $\SI{1.3(1)}{\milli\K}$ without measurable atom loss. We then apply a two-photon cooling step, in which we switch on the 1034-nm laser light, while simultaneously decreasing the 423-nm power to $P_{423} = \SI{340}{\micro\watt}$ ($I_{\text{b}} = 0.06I_{\text{sat,b}}$) and changing the 423-nm detuning. The 1034-nm laser beam has a beam waist of $\SI{1.2}{\milli\meter}$ and a power of $P_{1034} = \SI{5.7}{\milli\watt}$, resulting in an intensity of $I_{\text{ir}} = 100I_{\text{sat,ir}}$, with $I_{\text{sat,ir}} = \SI{0.46}{\milli\watt\per\centi\meter\squared}$ the saturation intensity of the 1034-nm transition. The 1034-nm laser is frequency stabilized to a cavity, the length of which is varied to tune $\Delta_{\text{ir}}$. We perform two-photon cooling for $t_{\text{tpc}} = \SI{10}{\milli\second}$, which we verified to be sufficient time for transient behaviours to dissapear. Finally, we record the cloud's size by performing absorption imaging after time-of-flight (TOF) expansion. We estimate the cloud temperature by measuring the cloud size for different time-of-flight expansion times. 

\section{IV. Results}
\setlabel{IV}{sec:experiment}

\begin{figure}
    \centering
	\includegraphics[width=\columnwidth]{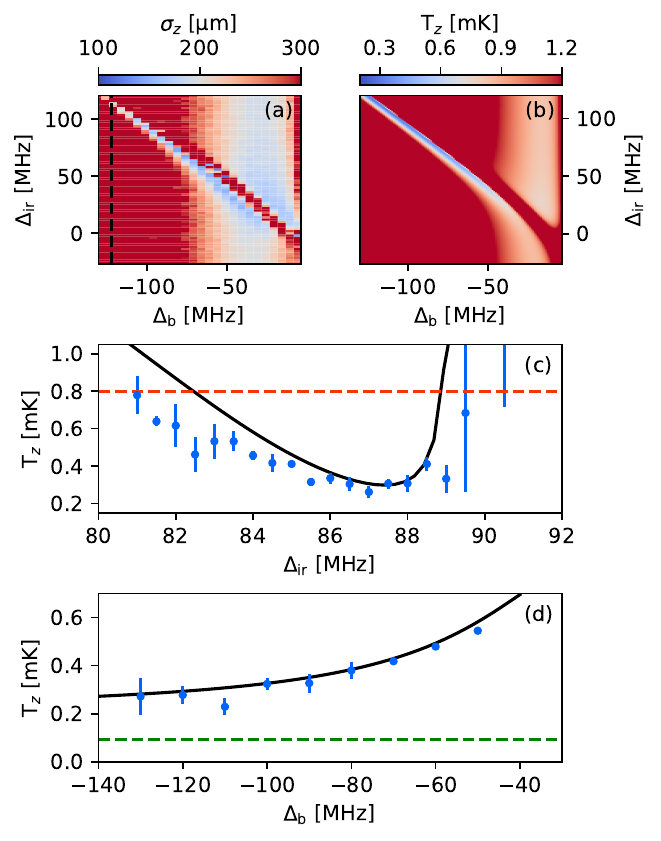}
	\caption {{\bf Sub-Doppler temperatures via two-photon cooling.} (a) Measured cloud's vertical width as a function of 423-nm detuning and 1034-nm detuning. Experimental parameters are: $P_{423} = \SI{340}{\micro\watt}$, $P_{1034} = \SI{5.7}{\milli\watt}$, $t_{\text{tpc}} = \SI{10}{\milli\second}$. (b) Simulated cloud's temperature for the parameters of (a). (c) Atomic cloud temperature measured along the vertical direction as a function of 1034-nm detuning (data points). The black line shows the theoretical calculation. The red dashed line represents the Doppler limit of the ${^1}P_1$ transition ($T_{\text{Doppler}} = \SI{0.8}{\milli\kelvin}$). (d) Lowest achieved temperature as a function of the 423-nm detuning (data points). The black line and the green dashed line show the theoretical prediction and the Doppler limit of the $4s5s~{^1}S_0$ transition ($T^{*}_{\text{Doppler}} = \SI{93}{\micro\kelvin}$), respectively. 
	}
	 \label{fig:2} 
\end{figure}

Using the experimental sequence described above, we perform the two-photon cooling for various detunings of the 1034-nm and 423-nm laser beams. The resulting measured cloud width (a proxy for temperature) and simulated temperatures for the experimental parameters are shown in Fig.~\ref{fig:2}(a, b). The plots display how the feature arising from Doppler cooling on the 423-nm transition is modified by the presence of the 1034-nm light, which leads to sub-Doppler temperatures along the two-photon resonance. In the two-photon cooling regime, the temperature decreases for larger detunings. We observe a temperature $T_z = \SI{ 260 \pm 30}{\micro\kelvin}$ for a detuning $\Delta_{\text{b}} = \SI{-122}{\mega\hertz} = - 3.5 \Gamma_{\text{b}}$ (dashed line in (a)). While the frequency of the 1034-nm light can be varied over a large range (up to \SI{1}{\giga\hertz}), the frequency range of the 423-nm light is limited by the AOM bandwidth. 

By varying the 1034-nm detuning for a fixed $\Delta_{\text{b}}$, the temperature decreases until reaching a minimum when approaching the two-photon resonance. \mbox{Figure~\ref{fig:2}(c)} shows the temperature measurement along the $z$-direction as well as the theoretical calculation, which corresponds to the parameters indicated by the black dashed vertical line in Fig.~\ref{fig:2}(a). The discrepancy between the data and the theoretical model can be attributed to uncertainty in the linewidth of the $4s5s~^{1}S_0$ state, and to the cloud's position relative to the laser beams and magnetic field null. The lowest temperature along the $x$-direction is about \SI{400}{\micro\kelvin}. The higher temperature compared to the other direction is likely due to the presence of a bias magnetic field at the cloud's position and small displacements of the cloud with respect to the primary cooling beams. 

The lowest achievable temperature depends strongly on the 423-nm detuning. By increasing the detuning and therefore reducing the scattering from the intermediate ${^1}P_1$ state, the temperature decreases in good agreement with theoretical predictions (see Fig.~\ref{fig:2}(d)). However, in our experiment, the largest usable detuning and lowest achievable temperature is limited by the scattering force arising from the MOT beams, which reduces for larger detunings, thereby causing atom losses. This limitations can be overcome by using this cooling scheme for atoms confined in optical potentials, where radiation pressure is not required for trapping. A more detailed analysis of the lowest achievable temperatures with the single 1034-nm beam configuration can be found in the Appendix\,\ref{AppendixB}.

The measurements shown in Fig.~\ref{fig:2} were performed by setting the polarization of the 1034-nm laser beam to be linear with an angle of 50$^\circ$ with respect to the $x$-axis. Due to the presence of magnetic sub-levels, our system involves five states, which makes the choice of the polarization crucial to the cooling efficiency. As the six counter-propagating 423-nm laser beams constitute our magneto-optical trap, their polarization is set to $\sigma_+$ and $\sigma_-$ for each pair. The overall change of the angular momentum of an electron between ${^1}S_0$ and $4s5s~{^1}S_0$ is zero. 

Conservation of angular momentum thus constrains the polarization of the 1034-nm laser beam. Specifically, the polarization must include a component that results in zero total angular momentum when combined with each of the 423-nm cooling beams. Therefore, when using a single dressing beam, the polarization of the 1034-nm beam must not be circular if it co-propagates with one of the MOT beams. If the 1034-nm beam is orthogonal to the MOT beam, its polarization must not be linear and parallel to the MOT beams.

\begin{figure}
    \centering
	\includegraphics[width=\columnwidth]{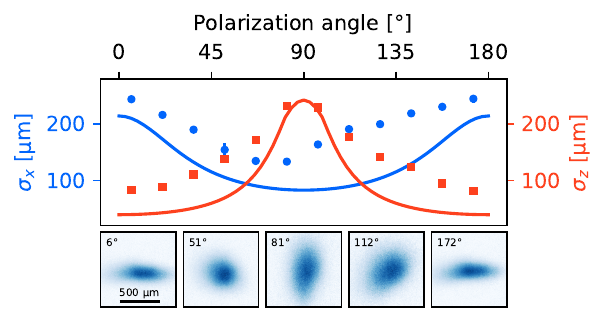}
	\caption {{\bf Polarization dependence.} Width of the atomic cloud along the horizontal (blue circles) and vertical (red squares) direction as a function of the polarization angle of the linearly polarized 1034-nm beam. Absorption images of the cloud are displayed underneath. At 0$^\circ$, the polarization points along the $x$-axis, at 90$^\circ$ along the $z$-axis (see Fig. \ref{fig:1}). This measurement was taken with the same experimental parameters of Fig. 2, except for $\Delta_{\text{b}} = -  \Gamma_{\text{b}}$. The solid lines correspond to theoretical simulations using the experimental parameters.
 }
	 \label{fig:3} 
\end{figure}

To verify this in the experiment, we rotate the linear polarization of the 1034-nm beam and record the cloud's width along the vertical and horizontal direction for different polarization angles. Figure~\ref{fig:3} shows the results. The blue points and the red squares represent the width along the horizontal and vertical direction, respectively. The bottom images show pictures of the atomic cloud, when applying a two-photon cooling procedure with varying linear polarization. Horizontal polarization (0$^\circ$) gives maximum cooling efficiency along the vertical direction, which is reflected in a horizontally elongated cloud and therefore a higher temperature along the horizontal direction. By rotating the polarization, the cooling efficiency decreases along the vertical direction and increases along the horizontal direction, reaching a maximum when the polarization is rotated by 90$^\circ$. At an angle of about 50$^\circ$, we obtain cooling along both directions, which results in temperatures well below the Doppler limit. 

The continuous lines in Fig.~\ref{fig:3} show the theoretical calculation for the corresponding MOT widths, which are smaller than the experimental results. Mismatch between theory and  experiment is likely due to a shift in the MOT position when transitioning from the pre- to two-photon cooling. The low-saturation regime of the ${^1}S_0 \rightarrow {^1}P_1$ transition leads to sensitivity to the imperfection and alignment of the MOT beams. The shift results in different perceived laser intensities and induces a bias magnetic field, which overall reduces the scattering rate and hence increases the MOT size. Retro-reflecting the 1034-nm laser beam and increasing the size of the MOT beams might help in reducing this shift.

The dressing beam does not introduce any additional loss channels to the MOT. The branching ratio between $4s5s~^1S_0 \rightarrow {^1}P_1$ and $4s5s~^1S_0 \rightarrow {^3}P_1$ is approximately 10'000:1 and thus the $4s5s~^1S_0$ state predominantly decays back into the cooling cycle. While the fraction of the population decaying to the short-lived $^3P_1$ state is not directly trapped, those atoms are not lost from the cooling cycle. Due its short lifetime of \SI{330}{\micro \second}~\cite{Mills2017efficient}, $^3P_1$ atoms are re-captured by the MOT beams after they have decayed back to the ground state. If this was not the case, the expected loss rate for 10\% population in $4s5s~^1S_0$ would be around \SI{50}{\per \second}, or \SI{20}{\milli \second}. The measured $1/e$ atom lifetimes are an order of magnitude larger, underlining that atoms decaying to the ${^3}P_1$ level are indeed recaptured by the MOT beams.  

\begin{figure}
    \centering
	\includegraphics[width=\columnwidth]{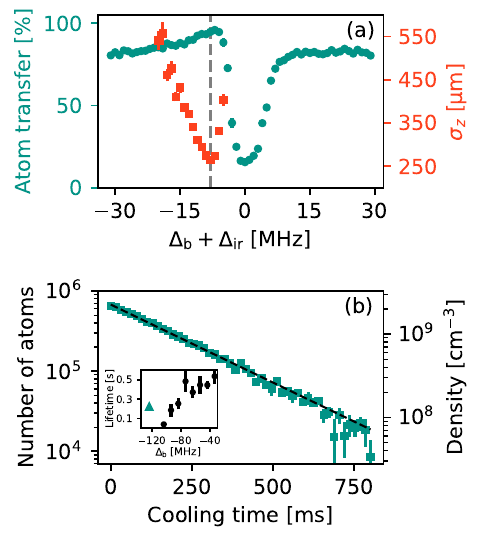}
	\caption {{\bf Transfer efficiency and lifetime.} (a) Atom transfer efficiency (teal circles) for a 10-ms long two-photon pulse as a function of the total detuning. The 423-nm laser detuning is set to $\Delta_b = -3.5\Gamma_{\text{b}}$, while we vary the frequency of the 1034-nm laser. As indicated by the size of the atomic cloud (red squares), the cloud is coldest at $\Delta_\text{b} + \Delta_\text{ir} =$ \SI{-8}{\mega \hertz} (dashed grey line). This minimum coincides with the maximum transfer efficiency of 96\%. (b) Number of atoms as a function of cooling time, measured at $\Delta_\text{b} + \Delta_\text{ir}=-8$ MHz -- grey dashed line  in (a). From the linear fit (dashed black line), we extract a $1/e$ lifetime of \SI{223(2)}{\milli\second}, marked by the teal triangle in the inset. The black circles in the inset show the lifetime of the MOT without two-photon cooling. The lifetime decreases as the blue detuning increases. We attribute this to a reduction in the scattering rate.
	}
	 \label{fig:4} 
\end{figure}

The two-photon cooling scheme does not require any technique to spectrally broaden the transition in order to improve transfer efficiency. After the pre-cooling step, the atomic cloud temperature is on the order of \SI{1}{\milli\kelvin}. This results in a Doppler broadening of $\Delta \omega_D=2 \pi \times \sqrt{4 \hbar \Gamma_{\text {blue }} \ln 2 /\left(m \lambda^2\right)} \simeq 2 \pi \times 2 ~\mathrm{MHz}$~\cite{Snigirev2019Fast}. For a typical MOT size of \SI{100}{\micro\meter} the atoms experience a Zeeman broadening of about \SI{3}{\mega\hertz}. The linewidth of the $4s5s~^1S_0$ state is similar to the Doppler and Zeeman broadening. Hence, nearly all atoms are transferred to the two-photon MOT. Figure~\ref{fig:4} shows the measured transfer efficiency of the pre-cooling atom number to the number of atoms left after 10 ms of two-photon cooling. The power and detuning of the 423-nm laser are kept constant at \SI{340}{\micro \watt} and $-3.5\Gamma_\text{b}$, respectively. At the two-photon resonance $\Delta_\text{b}+\Delta_\text{ir}=0$, the atom transfer efficiency has a minimum due to heating. As indicated by the size of the cloud, the minimum temperature is achieved roughly \SI{8}{\mega \hertz} red detuned (vertical dashed line) from the resonance. At this point, 96\% of atoms survive the cooling process. We measure the $1/e$ atom lifetime over a 800-ms long cooling pulse and find $\tau = $~\SI{223\pm 2}{\milli \second} (Fig.~\ref{fig:4}(b)). Consequently, the expected losses during 10 ms of cooling are roughly $1-e^{-10/223}=0.04$, which agrees with the measured transfer efficiency. Depending on the specific cooling parameters, the cloud reaches equilibrium after around 1-5 ms, implying that 98-99.5\% of loaded atoms are cooled down. 

The black points in the inset of Fig.~\ref{fig:4}(b) show the $1/e$ atom lifetime of the MOT without the 1034-nm beam on. The 423-nm laser power is set to the same value used for the two-photon cooling, $P_{\text{b}}$ = \SI{340}{\micro\watt}. At small detunings $\Delta_\text{b}$, the lifetime is limited by collisions with background gas. When detuning further red from the resonance, the lifetime decreases, which we attribute to a reduction in scattering rate of the 423-nm laser. In the two-photon MOT, the coupling to the $4s5s~^1S_0$ enhances the absorption of the 423-nm photons, increasing the confinement. Hence, for large $\Delta_\text{b}$, the observed lifetime of the two-photon MOT is longer than that of the single-photon MOT. This indicates that the lifetime is not limited by decay to dark states. 

\section{V. Conclusions}

We have reported experimental observation of sub-Doppler temperatures in a magneto-optical trap of calcium atoms. By using a single 1034-nm laser beam to implement a two-photon cooling scheme, we achieved temperatures as low as \SI{260}{\micro\kelvin}. The lowest achievable temperature is limited by atom losses due to the low scattering rate at large detunings. 
In optical potentials, which provide confinement, this method could achieve reach temperatures close to the Doppler limit of the upper transition ($T = \SI{93}{\micro\kelvin}$). We characterized the transfer efficiency from the standard MOT to the two-photon MOT, and showed an atom retention close to 100\%. This approach, which allows for a reduction of the cloud temperature while maintaining a straightforward implementation, represents an alternative to the standard sequence of two magneto-optical traps generally applied to alkaline-earth(-like) atoms. The two-photon scheme could be used as a bridge between the first and second stage MOT and aid reaching even higher phase space densities. The natural next step is to investigate the application of this two-photon cooling scheme for trapping calcium atoms in optical potentials.  


\section{Acknowledgments}

\begin{acknowledgments}
We thank Mark Saffman, Guido Pagano, and Ivo Knottnerus for insightful discussions on potential applications of this scheme. We extend our gratitude to Alexander Ferk for support on the experimental control system, and to Robin Oswald for assistance with the 423-nm polarization spectroscopy lock. This work was supported as a part of NCCR QSIT, a National Centre of Competence (or Excellence) in Research, funded by the Swiss National Science Foundation (grant number 51NF40-185902).  We acknowledge support from the Swiss National Science Foundation (SNF) under Grant No. 200021-227992. W.A. acknowledges the support from ETH Research Grant.
\end{acknowledgments}

\appendix
\renewcommand\thefigure{S\arabic{figure}}   
\setcounter{figure}{0} 
\renewcommand{\thesection}{\Alph{section}}%
\setcounter{section}{0}

\section{APPENDIX A: THEORETICAL CALCULATION OF TEMPERATURE}
\setlabel{A}{AppendixA}

Calculations of the forces and thus the temperatures achieved in our system follow similar calculations used in Ref.~\cite{Mehlstaeubler2008observation}. We calculate the light forces by summing up radiation pressure contributions from all lasers present in the system. In the low-saturation regime, the total force can be calculated by summing the radiation pressure contributions from each individual blue laser beam in the presence of the infrared beam, and the radiation pressure from the infrared beam in presence of the blue beams. The total force can be written as below:

\begin{equation}
\textbf{F} = \sum_i \left( \rho^{(i + \text{ir})}_{11} \hbar \textbf{k}_{\text{b}}^{(i)} \Gamma_{1} + \rho^{(i + \text{ir})}_{22} \hbar \textbf{k}_{\text{ir}} \Gamma_{2} \right),
\label{equation force}
\end{equation}

where the index $i$ runs over the blue laser beams. The terms $\rho_{11}$($\rho_{22}$) represent the sum of the diagonal elements of the internal steady-state density matrix for all magnetic sub-levels of the ${^1}P_1$ (4s5s${^1}S_0$) states calculated in the presence of the $i{^{th}}$ blue beam and the infrared beam. 

The dependence of this force on the velocity through the Doppler shift, leads to the friction coefficient $\alpha_k = -dF_k/dv | _{v=0}$ ($k = x,y,z$), which is the source of cooling. The stochastic nature of absorption and emission of photons results in a random walk in phase space, which is the source of heating. It is characterized by the diffusion coefficients $D_{\text{emiss}}$ and $D_{\text{abs}}$ for emission and absorption, respectively: 

\begin{equation}
\begin{aligned}
D_{\text{emiss}} = \sum_i \frac{1}{2} \frac{1}{3} &\left( \rho_{11}^{(i+\text{ir})} \left(\hbar k_\text{b} \right)^2 \Gamma_1 \right. \\  &\left.+ \rho_{22}^{(i+\text{ir})} \left(\hbar k_{\text{ir}} \right)^2 \Gamma_2 \right),
\label{equation diffusion emission}
\end{aligned}
\end{equation}

\begin{equation}
\begin{aligned}
\left( D_{\text{abs}} \right)_k  = \sum_i &\frac{1}{2} \left(   \rho^{(i+\text{ir})}_{11} \left(\hbar \textbf{k}^{(i)}_\text{b} \cdot \hat{\textbf{u}}_k \right)^2 \Gamma_1 \right. \\ & \left. + \rho^{(i+\text{ir})}_{22} \left(\hbar \textbf{k}^{(i)}_{\text{ir}} \cdot \hat{\textbf{u}}_k \right)^2 \Gamma_2 \right),
\label{equation diffusion absorption}
\end{aligned}
\end{equation}

where $\hat{\textbf{u}}_k$ is the unit vector along the $k$-axis. The quantities $\rho_{jj}^{(i+\text{ir})}$ are obtained by solving the Lindblad Master Equation in the presence of the $i^{th}$ blue laser beam and the infrared beam using the QuTiP package~\cite{Johansson2012qutip}. The balance between cooling and heating leads to the steady state temperature:
\begin{equation}
T_{k} = - \frac{D_{\text{emiss}}+\left( D_{\text{abs}} \right)_k}{\alpha_k}.
\label{equation temperature}
\end{equation}

\section{APPENDIX B: LOWEST ACHIEVABLE TEMPERATURE}
\setlabel{B}{AppendixB}

\begin{figure}
    \centering
	\includegraphics[width=\columnwidth]{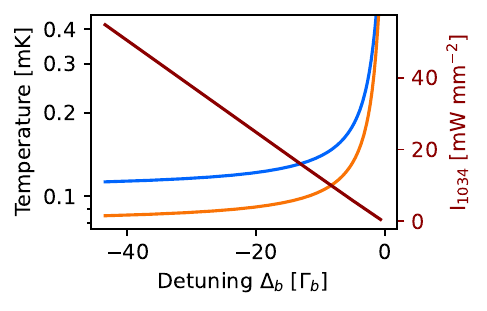}
	\caption {{\bf Theoretically calculated minimal temperatures.} Lowest achievable temperature was determined using a conjugate gradient optimization algorithm. The detuning and intensity of the 1034-nm beam are optimized for each 423-nm detuning. The intensity of the 423-nm light is constant and equal to \SI{2}{\milli\watt\per\centi\meter\squared}, and the polarization of the 1034-nm light is set to \SI{38}{\degree} relative to the $x$-axis. The blue line represents the optimal temperatures along the $x$- and $z$-axes, while the orange line shows the optimal temperature along the $y$-axis. The dark red line shows the corresponding intensity of the 1034-nm beam. For larger detunings, the intensity has to be increased to maintain the optimal cooling condition. 
	}
	 \label{fig:s1} 
\end{figure}

We simulated the minimum temperatures achievable in our system by running a conjugate gradient algorithm~\cite{Virtanen2020SciPy}. For each 423-nm detuning, we optimize the detuning and intensity of the \SI{1034}{\nano\meter} beam, at a fixed 423-nm light's intensity of \SI{2}{\milli\watt\per\centi\meter\squared}. Figure~\ref{fig:s1} shows the resulting optimal temperatures. The temperatures decrease rapidly within the first five linewidths. At large detunings, the temperatures asymptotically converge to \SI{85}{\micro\kelvin} along the $y$-axis and \SI{112}{\micro\kelvin} along the $x$- and $z$-axes. The temperature along the $y$-axis is below the Doppler limit of the upper ${^1}S_0$ state. This is due to the asymmetry of the Fano line shape~\cite{Magno2021Laser}. The optimal intensity of the 1034-nm beam increases with the 423-nm detuning.

The discrepancy between the optimal temperatures in the $x$-, $z$-, and \mbox{$y$- directions} arises because the 1034- nm beam is aligned along the $y$-axis, resulting in a polarization always close to the optimal value. In contrast, for the $x$- and $z$-axes, the polarization must be tuned to cool both directions simultaneously, resulting in a trade-off between the temperatures along the $x$- and $z$-axes (see discussion in the main text, and Fig~\ref{fig:3}). For our specific geometry, this optimal polarization is at \SI{38}{\degree} relative to the $x$-axis.

\bibliography{TwoPhotonBib}
\end{document}